\documentclass{tlp} 

\newcommand{\remove}[1]{}

\usepackage{caption}
\usepackage[fleqn,alignedleftspaceno]{amsmath}
\usepackage{amssymb}
\usepackage{multirow}
\usepackage{soul}
\usepackage{listings}
\usepackage{algorithm}
\usepackage{algpseudocode}
\usepackage{hyperref}
\usepackage[english]{babel}
\usepackage{biblatex}
\usepackage{csquotes}
\usepackage[linewidth=0.5pt]{mdframed}
\usepackage[frozencache,cachedir=.]{minted}
\usepackage{listings}
\usepackage{multicol}
\usepackage{newfloat}
\usepackage{wrapfig}
\usepackage{comment}

\usepackage{tikz}
\usepackage{pgfplots}
\usepackage{tikz-qtree}
\usetikzlibrary{calc,math,shapes.geometric,arrows,fit,matrix,automata,positioning,shapes.symbols,shapes.misc}
\usetikzlibrary{fit,decorations.pathmorphing,arrows.meta}
\usetikzlibrary{pgfplots.groupplots}
\usetikzlibrary{shadings,patterns}
\pgfplotsset{compat=newest}

\DeclareFloatingEnvironment{code}
\usepackage{url}
\newcommand{\step}[2]{s^{#2}_{#1}}
\newcommand{\locksynth}[0]{\textit{Locksynth }}

\newcommand{\mycomment}[1]{}
\urldef{\mailsa}\path|saratchandra.varanasi@ge.com, neerajm@utdallas.edu, gupta@utdallas.edu|

  \author[S.C. Varanasi, N. Mittal and G. Gupta]{Sarat Chandra Varanasi$^{1}$, Neeraj Mittal$^{2}$ and Gopal Gupta$^{2}$ \\        
       General Electric Research, NY, USA$^{1}$ \\
         The University of Texas at Dallas, TX, USA$^{2}$\\
         \email{saratchandra.varanasi@ge.com, \{neerajm, gupta\}\@utdallas.edu}}
\begin{document}

\title[Locksynth: Deriving Synchronization Code]{Locksynth: Deriving Synchronization Code for Concurrent Data Structures with ASP}

\maketitle

\begin{abstract}
We present \textit{Locksynth}, a tool that automatically derives synchronization needed for destructive updates to concurrent data structures that involve a constant number of shared heap memory write operations. \locksynth serves as the implementation of our prior work on deriving abstract synchronization code. 
 Designing concurrent data structures involves inferring correct synchronization code starting with a prior understanding of the sequential data structure's operations. Further, an understanding of shared memory model and the 
 synchronization primitives is also required. The reasoning involved transforming a sequential data structure into its concurrent version can be performed using Answer Set Programming and we mechanized our approach in previous work. The reasoning involves deduction and abduction that can be succinctly modeled in ASP. 
We assume that the abstract sequential code of the data structure's operations is provided, alongside axioms that describe concurrent behavior. This information is used to automatically derive concurrent code for that data structure, such as dictionary operations for linked lists and binary search trees that involve a constant number of destructive update operations. We also are able to infer the correct set of locks (but not code synthesis) for external height-balanced binary search trees that involve left/right tree rotations.  
Our tool \locksynth can systematically make the same judgments as a concurrency expert would, to generate correct synchronization code.\locksynth performs the analyses required to infer correct sets of locks and as a final step, also derives the C++ synchronization code for the synthesized data structures. We also provide a performance analysis of the C++ code synthesized by \locksynth with the hand-crafted versions available from the Synchrobench microbenchmark suite. To the best of our knowledge, our tool is the first to employ ASP as a backend reasoner to perform concurrent data structure synthesis.   
\end{abstract}


\section{Introduction}
\label{section:intro}
  We present our tool \locksynth which serves as an implementation and extension of techniques from our prior work on concurrent data structure synthesis \cite{varanasi2021iclp}. Given the background data structure theory and the knowledge of sequential data structure operations, \locksynth infers the correct set of locks for safe concurrent execution of the same data structure operations while also generating the C++ code, if the data structure dictionary operation involves a constant number of destructive updates to the heap. 
  We extensively utilize Answer Set Programming to perform the back-end reasoning that facilitates the concurrency synthesis. We first present the technique behind \locksynth and then describe C++ synthesis on an External Binary search tree example. For the C++ versions synthesized, we have also compared their performance against equivalent hand-crafted versions (if available) from the Synchrobench microbenchmark suite \cite{gramoli2015more}. It is well known that the design of concurrent programs is notoriously difficult; they are also difficult to debug. More often, concurrent programs are carefully designed and a proof of correctness is provided manually \cite{vafeiadis2006proving, herlihy2020art}. Bounded model checking is another technique used to find error traces in concurrent programs \cite{emerson2000reducing, vechev2008deriving} and few works have also verified concurrent data structures in a semi-automated fashion \cite{vechev2010abstraction}. Compositional program verification  approaches for concurrency perform Rely-Guarantee reasoning \cite{vafeiadis2006proving} and have also mechanized their proof procedure by specifying abstract domains of linked lists represented in separation logic \cite{distefano2006tacas, vafeaidisthesis}. Our approach using \locksynth relies entirely on encoding the knowledge and reasoning performed by a concurrency expert into ASP and making that knowledge executable. 
  We do not take generated concurrent code and verify its correctness, rather use a lot of domain knowledge about the data structure---its background theory, knowledge of data structure precondition, and destructive pointer linkage operations---and perform reasoning to infer the right set of locks to acquire in a safe concurrent execution. \locksynth employs deductive and abductive reasoning approaches that are widely used in several Artificial Intelligence and Knowledge Representation \& Reasoning problems and domains. Moreover, the effects of destructive updates are modeled in ASP similar to how many action languages in AI are encoded in ASP \cite{gelfond2014knowledge}. We assume familiarity with Answer Set Programming; a comprehensive background on ASP can be found elsewhere \cite{gelfond2014knowledge}. \locksynth itself is implemented in SWI-Prolog and its backend reasoning engine uses the Clingo ASP solver\cite{gebser2018potsdam}. Our main contributions are the following:
  \begin{enumerate}
      \item Development of the \locksynth tool implemented completely in SWI-Prolog that uses ASP paradigm for its backend reasoning engine. 
      \item Ability to generate the C++ synchronization code based on the locks inferred by \locksynth when given the correctly annotated traversal code
      \item Benchmarking of the synthesized code versions with the hand-crafted versions from the Synchrobench microbenchmark suite.
  \end{enumerate}
  \par
  The rest of the paper is organized as follows: Section \ref{section:model} describes the problem statement for concurrency synthesis performed in this paper and the system model assumed for concurrent execution. Section \ref{section:idea} motivates the general idea of using AI to perform concurrency synthesis and concludes with a general procedure for lock synthesis. Section \ref{section:locksynth} describes in detail the input to \locksynth and the details of its reasoning procedure while using Linked Lists as a running example. Section \ref{section:cpp} describes the actual C++ synthesis after inferring the correct set of locks and the assumptions involved. Section \ref{section:cpp} uses the External Binary Search tree insert operation as its running example. Section \ref{section:synchrobench} performs a comparison of the synthesized code versus the hand-crafted concurrent versions from the Synchrobench microbenchmark suite. Section \ref{section:related} describes related work done in the field of concurrent data structure synthesis followed by conclusion and future work in Section \ref{section:conclusion}

 \section{Problem Statement and System Model}
 \label{section:model}
Our goal is to derive a safe concurrent algorithm when given a set of sequential programs, along with their appropriate preconditions and postconditions. The sequential programs are expressed in an abstract imperative language containing only destructive update steps and \textit{if} conditional checks.   
Each destructive update step modifies the contents of a heap cell atomically. The heap cell represents a node structure common in languages such as C and Java. Heap cells are part of shared memory, which allows concurrent accesses and modifications.  
The shared memory itself is sequentially consistent. Sequential consistency implies that any atomic destructive update step performed on the heap is immediately visible to every other thread in the system accessing the heap. 
We assume that \textit{if} conditional checks may be performed non-atomically. The expressions within the conditional checks are written in a simple assertion language and allow for only conjunctions of assertions. 
  Each assertion borrows its semantics (truth value) from the background theory of the data structure.  Any disjunctions present in the preconditions can be broken up into multiple cases.  Every data structure operation is essentially a straight-line program
  %
  performing destructive updates on a fragment of the data structure, residing in shared memory. Restriction of straight-line program does not affect the synthesis for the data structures considered, however, it limits the kinds of operations that \locksynth can synthesize. For example, linked list reversal involves modification of heap cells proportional to the length of the list and is not supported. Hence, the number of destructive updates is constant.  However, for other non-trivial operations such as left/right rotations for balanced binary trees, our approach is sufficient.  Note that the background theory is manually provided as input by the user. 
  
\smallskip\noindent\textbf{Assumptions on Concurrent Execution:}
We assume that the concurrent threads destructively updating the heap interleave in a certain order. The only synchronization primitive available is the \textit{lock} statement. Locks can be acquired on the heap cells. 
 
\smallskip\noindent\textbf{Requirements of a Correct Concurrent Algorithm:}
  We impose certain requirements on a correct concurrent algorithm. Our technique synthesizes only algorithms satisfying these requirements. A correct concurrent algorithm that modifies a heap (shared memory) should satisfy the following requirements: 
  \begin{enumerate}
      \item[R1] Every thread must acquire locks on the heap cells it is going to modify. 
      \item[R2]  Every thread must validate its precondition after lock acquisitions and before performing destructive update step. If the validation fails post lock acquisition, the respective thread should relinquish its locks without performing its destructive updates. 
      \item[R3] Every thread must acquire locks on the heap cells present in the precondition of its data structure operation.
      \item[R4] Every thread shall acquire and release locks in a uniform order on the heap cells
  \end{enumerate}
  Requirement R1 refers to each thread performing destructive updates in isolation. Requirement R2 is necessary for fine-grained concurrent data structures. Because lock acquisition itself is not an atomic step, a data structure may be modified during the time it takes to acquire the requisite locks. Hence a post lock-acquisition validation is necessary. Requirement R3 is not immediately obvious but is relevant in the context of concurrent data structures described in this paper. Requirement R4 is needed for deadlock avoidance.

\section{Motivation and General Idea}
\label{section:idea}
Human experts, when designing concurrent data structures, have a firm understanding of the data structure representation and the sequential behavior of the dictionary operations supporting the data structure. To transform a given data structure comprised of sequential dictionary operations, the human expert deduces how the operations interact concurrently. The expert checks the concurrent interaction on some sample instances of the data structure.  The sample instances are selected to allow maximum concurrent interaction of the data structure operations. In the presence of maximum possible concurrent interaction, the expert deduces the right set of locks to be acquired for each data structure operation. They know that the locks acquired are adequate by giving a lock adequacy argument. Our tool executes and realizes this reasoning done by a human expert in ASP. We next describe our general technique. \par\noindent 
  
We only consider data structure operations that destructively modify a given data structure. That is, we completely ignore the part of the data structure operations that involves traversal. In order to perform actual code synthesis, we also assume that the traversal code is given to us. It is assumed that the data structure's destructive update operations are only comprised of a sequence of atomic write steps.   
Let the data structure $D$ be supported by dictionary operations $\{\sigma_1, \sigma_2, \ldots, \sigma_n\}$. The only changes that can be made to the data structure are attributes of nodes on the heap. Let $l(\sigma_i)$ denote the number of write steps involved in $\sigma_i$. Then, each $\sigma_i$ is a sequence of atomic write steps: $(\step{\sigma_{i}}{1},~\step{\sigma_{i}}{2},~\dots,~\step{\sigma_{i}}{l(\sigma_{i})})$. Let $pre(\sigma_i)$ denote the precondition for $\sigma_i$. Let $locks(pre(\sigma_i))$ denote the set of locks guessed based on the given precondition $pre(\sigma_i)$. \par\noindent 
  \par\noindent
    
The set of locks guessed can be based upon any heuristic that a reasoner thinks is adequate in the presence of concurrent interactions. Because locks are acquired on individual nodes of the data structure, every lock in $locks(pre(\sigma_{i}))$ is associated with a node. By a slight abuse of notation, let $\delta \in D$ mean that $\delta$ is some instance of the data structure $D$. We call some $\delta_m \in D$ as maximally applicable if each  $pre(\sigma_i)$ is applicable to $\delta$. Intuitively, maximally applicable instances capture all possible interactions of $\sigma_1, \ldots, \sigma_n$. For the data structures discussed in this paper, we are able to find data structure instances satisfying all the preconditions for $\sigma_1$ through $\sigma_n$. Further, we require that all the preconditions to be applicable on the instance. This is important because, after lock acquisition, we want to check if the application of the data structure operations indeed satisfy invariants associated with the data structure, without which we cannot guarantee that the application of the operations leaves the data structure in a consistent state. 


To illustrate, consider $LinkedList$ supported by $\{insert, delete\}$. Every instance of $LinkedList$ has two sentinel nodes $h$ and $t$ at the beginning and at the end respectively. Every other node sits between $h$ and $t$. Let the target node (key) to be inserted (deleted) from the list be denoted by $\tau$. The relation $edge(x,y)$ signifies the node $y$ is the next pointer of node $x$. The notation $k_x, k_y$ is used to denote the key values of nodes $x, y$ respectively. The relation $reach(x)$ signifies that the node $x$ is reachable from head of the list. 
  With these notions, preconditions $pre(insert)$ and $pre(delete)$ are defined as follows: \par\noindent 
 $pre(insert(\tau)) \equiv \exists \tau, x, y  \ \neg reach(\tau) \land reach(x) \land edge(x,y) \land k_x < k_{\tau} < k_y $ \\
$pre(delete(\tau)) \equiv \exists \tau, x, y \ reach(\tau) \land reach(x) \land edge(x,\tau) \land edge(\tau, y) $ \\
A sample Linked List is $\delta = \{edge(h, t)\}$ with $\{k_h < k_t\}$. However, $\delta$ is not maximally applicable as the delete operation is not applicable. But the list $\delta' = \{edge(h,x), edge(x,t)\}$ with $k_h < k_x < k_t$ is maximally applicable as both insert and delete are applicable. The program specification in Hoare-triples style and its equivalent concurrent version for Linked List Insert and Delete operation can be found in our prior work \cite{varanasi2021iclp}.  Note that, the sentinel node approach to linked list definition subsumes the corner cases of inserting or deleting an element at the beginning or end of the list. However, in an alternative encoding of linked lists that does not take sentinel nodes into account, separate preconditions and program steps for the corner cases must be specified. 
The steps involved to generate synchronization code for $\sigma_i$ are  outlined in the following algorithm. By a coarse-grained locking scheme, we mean a naive locking algorithm that associates a single lock for the whole data structure. A coarse-grained locking scheme is inefficient when compared to a fine-grained locking scheme that associates locks with individual heap cells of the data structure. Note that, this synchronization code is combined with data structure traversal code as shown in Section \ref{section:cpp} when performing actual C++ code synthesis.

\parbox{7cm}{
\begin{algorithm}[H]
\begin{algorithmic}
 \Procedure{GenerateSynchronizationCode}{$\sigma_i$}
\State Guess set of locks, represented by $locks(pre(\sigma_i))$
\State Check if, any of $\sigma_1, \sigma_2, \ldots, \sigma_n$, when applied on maximal instance $\delta_{m}$, falsifies $pre(\sigma_i)$
\If{ $pre(\sigma_i)$ is not violated due to interference by $\sigma_1, \sigma_2, \ldots, \sigma_n$} 
   \State  Use fined grained-locking by acquiring locks on every node in $locks(pre(\sigma_i))$
\Else
    \State $locks(pre(\sigma_i$)) is inadequate
    \State  Use Coarse-grained locking
\EndIf
\EndProcedure
\end{algorithmic}
\caption{Algorithm to Generate Synchronization Code for Operation $\sigma_i$}
\label{alg:general_algo}
\end{algorithm}
}

If the locks guessed for $\sigma_i$ are adequate, the concurrent code generated would look as shown in Listing \ref{code:template_code}. In the generated concurrent code, $\{x_1, x_2, \ldots, x_k\}$ constitute the set of nodes guessed as part of $locks(pre(\sigma_i))$, to be locked for a safe concurrent execution. 
Note that the extra validation step is necessary as the data structure might have changed by the time the correct locks were acquired. This is exactly the kind of programs we mentioned in \textit{Requirement R2}.  If the validation step fails then the locks are relinquished without any modification. If \locksynth cannot find the correct set of fine-grained locks, it uses a simple coarse-grained locking scheme. Further, the lock $x_1, x_2, \ldots, x_k$ are acquired and released in a uniform order. For the tree-based data structures considered in this paper, locks on nodes that appear earlier in the preorder traversal of the tree from the root are acquired and released before the locks on nodes that appear later in the same preorder traversal of the same tree from the root.

\begin{code}
    \begin{minted}[frame=single, escapeinside=||,mathescape=true,fontsize=\scriptsize]{c}
  lock(|$x_1$|);  lock(|$x_2$|);  |\ldots|; lock(|$x_k$|);
  if(validate(|$pre(\sigma_i)$|){ 
     |$\step{\sigma_{i}}{1}$| ; |$\step{\sigma_{i}}{2}$|; |\ldots|; |$\step{\sigma_{i}}{l(\sigma_{i})}$|;
  }
  unlock(|$x_1$|); unlock(|$x_2$|); |\ldots|; unlock(|$x_k$|)
\end{minted}
\captionof{listing}{\bf Synchronization Code  generated for operation $\sigma_i$}
\label{code:template_code}
\end{code}


We mentioned that the  set $locks(pre(\sigma_i))$ is based upon some heuristic. Currently, the heuristic is based upon locking every node cited in the precondition. We designate this heuristic as \textit{precondition locking strategy}. This strategy works for the data structures considered in this paper which are Linked Lists, External BSTs and External Balanced BSTs. This is exactly as stated in Requirement \textit{R3} in our assumptions. 
In general, a guessed set of locks can fail the lock adequacy check.
For instance consider locking only node $x$ in $pre(insert(\tau))$ for a linked list. In such a case, before an insert operation is performed in the window $\{x, y\}$, the node $y$ may be removed if it is not locked. This results in an incorrect insertion of the target node $\tau$.
Note that $R3$ works in conjunction with maximal instance definition used in this paper. Relaxing $R3$ will require stronger restrictions on maximal instance construction, which is outside the scope of this paper.  
We next describe how \locksynth is used to transform a sequential data structure into a concurrent one using several reasoning tasks. They are detailed in the subsequent sections. We first start with the input that \locksynth takes. 

\section{Input to \locksynth and its reasoning procedure}
\label{section:locksynth}
\begin{enumerate}
\item The first component is \textit{background theory} of the data structure. The background theory contains the data structure definition and the invariants associated with the data structure. It also identifies the properties of the data structure that are time-dependent (fluents). The background theory is encoded in ASP. A sample background theory for Linked Lists is presented in Listing \ref{code:background_theory}. 

\item The second component is the \textit{sequential data structure knowledge}.  The sequential data structure knowledge consists of the dictionary operations along with the preconditions under which they can be executed. The sequence of atomic write steps involved in each of the dictionary operations are also listed. (Listing \ref{code:seq_ds_knowledge}). The sequential data structure knowledge is written in Prolog and translated into ASP code when performing appropriate reasoning tasks by \locksynth.
\end{enumerate}
\begin{code}
\begin{minted}[frame=single,fontsize=\scriptsize]{prolog}
% Representation of List
list :- edge(h, X), key(h, KH), key(X, KX), KH < KX, suffix(X).
suffix(X) :- edge(X, Y), key(X, KX), key(Y, KY), KX < KY, suffix(Y).
suffix(t). 
% Abstractions in Linked List Theory
reach(h).      reach(X) :- edge(Y, X), reach(Y).
present(K) :- reach(X), key(X, K).
% Invariants and Fluents
invariant(list). fluent(list). fluent(reach). fluent(edge). 
fluent(suffix). fluent(present). 
\end{minted}
\captionof{listing}{\bf Sample Background Data Structure Theory}
\label{code:background_theory}
\end{code}

\begin{code}
\begin{minted}[frame=single,fontsize=\scriptsize]{prolog}
operation(insert).        operation(delete). % Operations: insert, delete
atomic_step(link). % assigning pointers is the only atomic step
modifies(link(x,y), x). % Similar to x.next := y where x is modified
causes(link(x,y), edge(x,y)). % Causal meaning of x.next := y; link operation creates an edge

% Precondition, program steps and post condition for insert 
precondition(insert,  [reach(x),edge(x,y),not(reach(target)), lt(kx, ktarget), lt(ktarget, ky]).
% lt is the numerical less than relation
program_steps(insert, case1, [link(x, target), link(target, y)]).
postcondition(insert,  [reach(target)]).

% Precondition, program steps and post condition for delete 
precondition(delete, [reach(x), edge(x, target), edge(target, y)]).
program_steps(delete,  [link(x, y)]).
postcondition(delete,  [not(reach(target))]).
\end{minted}
\captionof{listing}{\bf Sample Sequential Data Structure Knowledge}
\label{code:seq_ds_knowledge}
\end{code}
 The \locksynth procedure uses the input provided by the user and performs the requisite reasoning in ASP. The top-level procedure called {\tt synth\_concurrent\_code}, implemented in Prolog, is shown in Table \ref{tab:locksynth_procedure}. The input provided by the user---sequential data structure knowledge and background theories---is processed by another Prolog procedure that
 prepares it for use by the {\tt synth\_concurrent\_code} procedure. The procedure for preparing data and generate final code text (\texttt{generate\_code}) is not shown here but can be found in \textit{Locksynth}'s GitHub repository\footnote{\url{https://github.com/sarat-chandra-varanasi/locksynth/blob/main/lsynth.zip}}.
 Procedure {\tt synth\_concurrent\_code} concretizes Algorithm \ref{alg:general_algo}; each of its steps uses the relevant pieces of knowledge from the input. Each step either generates code, transforms code, or checks that the generated code satisfies some condition. These steps are explained next.
 \begin{table}
 \begin{tabular}{p{6cm}p{6cm}}
 \begin{minted}[frame=single,fontsize=\scriptsize,escapeinside=||]{Prolog}
synth_concurrent_code(Op, Code) :-
  %% ASP Reasoning: Listing |\ref{code:max_instance}|
  pick_maximal_instance(Instance), 
  precondition(Op, Pre),
  program_steps(Op, Steps),
  %% ASP Reasoning: Listing |\ref{code:interfere_abducibles}|
  generate_interference_rules(Rules),  
  guess_locks(Pre, Locks), 
  findall(Inv, invariant(Inv), Invariants), 
  %% ASP Reasoning: :Listing |\ref{code:program_order}|
  check_program_order(Instance, Steps,
     Invariants, Order), 
  %% ASP Reasoning: Listing |ref{code:locking}|
   check_locks_adequate(Instance, Locks),  
  generate_code(Rules,Locks,Pre,Order,Code).
 \end{minted}
 &
 \begin{minted}[frame=single,fontsize=\scriptsize,escapeinside=||]{Prolog}
 Bindings for Linked List Insert Operation
 ------------------------------------------
 Op = insert
 Pre = [reach(x),edge(x,y),kx<ktarget, 
        ktarget<ky]
 Instance = [edge(h,x),edge(x,t)]
 Rules = <List of rules as per Listing 5>
 Invariants = [:- not list]
 Steps = [link(x,target), link(target,y)]
 Order = [link(target, y), link(x, target)]
 Code = [lock(x),lock(y),
         if(validate(Pre)),
            link(target,y),
            link(x,target),
         unlock(x), unlock(y)]
 
 \end{minted}
 \end{tabular}
  \captionof{table}{Synthesis Procedure (left) \& Prolog Bindings for Linked List Insert (right)}
\label{tab:locksynth_procedure}
 \end{table}
 
 \smallskip\noindent{\bf Automatically Selecting Maximally Applicable Instance:}
 Selecting a maximally applicable instance is a constraint satisfaction problem.\locksynth generates candidate lists (trees) to check if they are maximally applicable. For example, given an instance $\{edge(h,x),edge(x,t)\}$ for linked list, the ASP program to check maximally applicable instance is found in Listing \ref{code:max_instance}.

\smallskip\noindent{\bf Concurrent Interaction Modelled in ASP:}  
 Concurrent interaction of operations is succinctly modeled in ASP. We state that an operation may happen or not happen. This generates $2^c$ possible worlds in which $c$ different operations may occur or not occur. Note that $c$ is constant w.r.t a data structure and is usually small. This  concurrent interaction is modeled by the use of abductive reasoning in ASP. \locksynth abduces whether interference from an operation may or may not happen. 

 \begin{code}
    \begin{minted}[frame=single,fontsize=\scriptsize]{prolog}
Program Fragment to Check if edge(h,x). edge(x,t) is maximally applicable:
---------------------------------------------------------------------------
 edge(h,x). edge(x,t). key(h,kh). key(x,kx). key(t,kt). lt(kh, kx). lt(kx, kt).
 lt(kh, kt).
 :- not precondition_insert.  :- not precondition_delete.
 Output model (SAT):  edge(h,x). edge(x,t). key(h, kh). key(x,kx). key(t,kt). lt(kh,kx). lt(kx, kt). 
 lt(kh,kt).  precondition_insert. precondition_delete.
     \end{minted}
     \captionof{listing}{\bf Maximally Applicable instance viewed as  Constraint Satisfaction}
     \label{code:max_instance}
     \end{code}
 Further, \locksynth also combines rules from background theory and sequential data structure knowledge to generate an ASP program. This ASP program is used to infer the effects of concurrent interactions on the maximal instance. The maximal instance itself is added to the ASP program as a set of facts. The ASP program modeling concurrent interactions should not only take into account interference but also consider the invariants associated with the data structure and evolution of state of the heap.
 Invariants are directly mapped to ASP constraints, fluents are reified into temporal domain, and any potential concurrent interference by each operation is modeled by an \texttt{interfere} predicate. The notions of fluents, commonsense law of inertia are standard idioms used in the Planning domain by the AI community \cite{gelfond2014knowledge}. 
 For example, in case of linked lists, interfere by insert or delete operation is modeled by rules in Listing \ref{code:interfere_abducibles}. Even loops over negation (ELON) capture abduction in ASP \cite{gelfond2014knowledge}. 
 The predicates {\tt interfere(insert(X,Target,Y))} and {\tt interfere(delete(X,Target,Y))} are part of even loops over negation. 
 \begin{code}
\begin{minted}[escapeinside=||,mathescape=true,frame=single,fontsize=\scriptsize]{prolog}
Program Fragment to model concurrent interaction:
------------------------------------------------
% interfere(insert) may or may not happen
interfere(insert(X,Target,Y), T) :- 
 precondition_insert(X,Target,Y,T),not neg_interfere(insert(X,Target,Y),T)).
neg_interfere(insert(X,Target,Y),T) :-
 precondition_insert(X,Target,Y,T), not interfere(insert(X,Target,Y,T)).
% interfere(delete) may or may not happen 
interfere(delete(X, Target,Y), T) :- 
 precondition_delete(X,Target,Y,T),not neg_interfere(delete(X,Target,Y),T)).
neg_interfere(delete(X,Target,Y),T) :-
 precondition_delete(X,Target,Y,T),not interfere(delete(X,Target,Y),T)).
 Output Model: 2^c possibilities of interfere insert and interfere delete
 ------------
\end{minted}
\captionof{listing}{\textbf{Rules that consider all possible concurrent interactions}}
\label{code:interfere_abducibles}
\end{code}
    
The rewrite of fluents and invariants using time argument, along with maximal instance added as facts. is shown in Listing \ref{code:reified_program}. Time domain ranges from 0 to a maximum value.  
{
\scriptsize
\begin{code}
\begin{minted}[frame=single,fontsize=\scriptsize]{prolog}
% Background theory fluents reified into time
list(T) :- time(T), edge(h,X,T), key(h,KH), key(X,KX), KH < KX, suffix(X,T).
suffix(X, T) :- time(T),edge(X,Y,T),key(X,KX),key(Y,KX), KX < KY,suffix(Y,T).
suffix(t, T) :- time(T). 
reach(h, T) :- time(T).  reach(X, T) :- time(T), edge(Y, X, T), reach(Y, T).
present(K, T) :- time(T), reach(X, T), key(X, K).
% Invariant mapped to constraint
:- time(T), not list(T).
% Maximal instance added as facts
edge(h, x, 0). edge(x, t, 0). key(h, kh). key(x, kx).
key(y, ky). <(kh, kx). <(kx, ky).
% Consequences of interference
edge(X,Target,T+1) :- interfere(insert(X,Target,Y), T).
edge(Target,Y,T+1) :- interfere(insert(X,Target,Y), T).
edge(X,Y,T+1) :- interfere(delete(X,Target,Y), T).
\end{minted}
\captionof{listing}{\textbf{Background theory projected into temporal domain}}
\label{code:reified_program}
\end{code}
}
Finally, \locksynth also uses sequential data structure knowledge to generate consequences of \texttt{interfere\_insert} and \texttt{interfere\_delete}. 
All the above rules shown in Listings \ref{code:interfere_abducibles} and \ref{code:reified_program} enable \locksynth to make judgments by the several ways the data structure operations can concurrently interfere. 

\smallskip\noindent{\bf Program Order Reasoning in ASP:}  Corresponds to \texttt{check\_program\_order} in \textit{Locksynth} procedure.
 Order of the program steps is important in a concurrent execution as an incorrect order can break an invariant. For instance, if the program steps of linked list insert are executed in the order:\texttt{link(x,target)}, followed by \texttt{link(target,y)}, then, the invariant \texttt{list} would be broken. However, by changing the order of execution to: \texttt{link(y, target)}, followed by \texttt{link(x, target)}, would not break the invariant.  \locksynth permutes the order of the program steps and finds the right sequence of the data structure program steps that does not break invariants. 
 If no such program order exists, then \locksynth falls-back to a coarse-grained locking scheme.  \locksynth allows a maximum of $l(\sigma_i) + 1$ time steps to check for any invariant violation from time 0 to $l(\sigma_i)$. 
 The rules for Linked List are shown in Listing \ref{code:program_order}.
\begin{code}
 \begin{minted}[frame=single, escapeinside=||, mathescape=true,fontsize=\scriptsize]{prolog}
   Program Fragment that checks for correct order of insert operation steps:
   --------------------------------------------------------------------------
  time(0..2) % |$l(insert) = 2$| 
  precondition_insert(X,Target,Y,0) :- reach(X,0),edge(X,Y, 0),key(X,KX),key(Target,Ktarget), 
     key(Y, KY), KX < KTarget, KTarget < KY.
  link(x, target, 0). % Incorrect order
  link(target, y, 1). % Incorrect order 
  % rule for edge creation
  edge(X,Y,T+1) :- link(X,Y,T), not modified(X,T).
  modified(X,T) :- link(X,Y,T).  
  :- time(T), not list(T). % Check for invariant violation
  Output Model (UNSAT) :   False % Correct order is: link(Target, Y, 0). link(X, Target, 1)
 \end{minted}
 \captionof{listing}{\bf Program Order Checking}
 \label{code:program_order}
 \end{code}
 To perform this task, \locksynth combines Listings  \ref{code:reified_program} and \ref{code:program_order} 
 and checks for satisfiability. If the union of two programs is unsatisfiable, then that permutation must be rejected. If their union is satisfiable, then selected order is adequate and can be accepted. The pointer link operations have their effects captured as transition rules common in Action languages ASP encoding with commonsense rules of inertia.  The permutation of program order is given by \locksynth and fed as facts to ASP. However they can be modeled directly in ASP using even loops but the corresponding ASP solving would take more time and require more executability conditions.   
 
\smallskip\noindent{\bf Lock Adequacy Reasoning in ASP:} Corresponds to \texttt{check\_locks\_adequate} in \locksynth procedure.
Checking Lock Adequacy involves looking for any precondition violation in the presence of interference and locking heuristic. Lock Adequacy check uses the same interference model (on maximal instance), but ensures that interference too obeys rules of locking.
That is, the same assumptions that interference may or may not occur are asserted as before. However, now, the consequences of interference are pre-empted if there is a lock involved. 
For instance, a pointer from $x \ to \ y$ cannot be changed unless the operation acquires a lock on $x$. Therefore, the consequences of interference from Listing \ref{code:reified_program} are re-written, to account for locking, shown in Listing \ref{code:locking}.
\begin{code}
\begin{minted}[frame=single,fontsize=\scriptsize]{prolog}
Program Fragment that checks for lock adequacy
-----------------------------------------------
edge(X,Target,T+1):- interfere(insert(X,Target,Y),T),not locked(X),not locked(Y).
edge(Target,Y,T+1):-interfere(insert(X,Target,Y),T),not locked(X),not locked(Y).
edge(X,Y,T+1):-interfere(delete(X,Target,Y),T),not locked(X),not locked(Target).
locked(X) :- precondition_insert(X, Target, Y).
locked(Y) :- precondition_insert(X, Target, Y).
falsify :- precondition_insert(X,Target,Y,T), not precondition_insert(X,Target,Y,T+1).
:- not falsify 
Output Model : False (UNSAT) %% Unsat implies locks are adequate!
\end{minted}
\captionof{listing}{\textbf{Checking lock adequacy}}
\label{code:locking}
\end{code}
Consider Listing \ref{code:locking}, where we are trying to check if there is any trace behavior of concurrent operations that can break the precondition for a particular insert operation. We acquire the 
locks for based on the insert operation corresponding to \texttt{precondition\_insert}. We now check if there is any interference via insert or delete that can break \texttt{precondition\_insert}. We should not arbitrarily modify the data structure and break \texttt{precondition\_insert}, rather allow only interference on nodes that are unlocked. If the window of modification of an interference operation intersects with the window corresponding to \texttt{precondition\_insert}, and the nodes involved in the window are locked (according to the locking heuristic), then the interference operation should be pre-empted. If the window intersects and the nodes are left unlocked, then the interference operation can destructively update the data structure. This destructive modification essentially captures another operation getting ahead of the current insert operation (corresponding to \texttt{precondition\_insert}) in a concurrent execution that could potentially break invariants or preconditions and would leave the data structure in an inconsistent state. Note that Listing \ref{code:locking}, captures \locksynth performing lock adequacy reasoning at a particular time instance in the concurrency behavior modeling, where locks are acquired and any interference that can potential wreak havoc is allowed to modify the data structure. This procedure will be repeated till the maximum time of the number of interference operations applicable on the maximal instance based upon different key orderings of target keys.    

To check lock adequacy for given operation, \locksynth combines listings \ref{code:interfere_abducibles}, \ref{code:reified_program} and \ref{code:locking} and checks for falsifiability of \texttt{precondition\_insert}. If \texttt{precondition\_insert} is falsified, then the locks are inadequate. If \texttt{precondition\_insert} is not falsified at any point in the trace, then the locks are adequate and internally, call to \texttt{check\_locks\_adequate(insert,\_,\_,\_)} would succeed. Note that the predicate \texttt{falsify} represents the notion whether precondition insert is falsified. If the program is satisfiable (produces a model), then \texttt{falsify} is necessarily present in the model (due to constraint \texttt{:- not falsify}).  That is, the guessed locks are inadequate. However, if the program is unsatisfiable, then no concurrent operation can break the precondition for insert (\texttt{falsify} cannot be in any model). That is, the guessed locks are adequate. 
\par To illustrate a case where an inadequate number of locks being acquired will result in a lock adequacy failure, consider the delete operation where a lock is only acquired on the predecessor to the target node. Assume the instance is $\{edge(x,\tau), edge(\tau,y)\}$ and only lock on $\{x\}$ is acquired. Then, a concurrent interference by an insert operation can change the heap to $\{edge(x,\tau), edge(\tau,\tau'), edge(\tau',y)\}$, where $\tau'$ is the target node for a concurrent insert operation whose window is $\{\tau, y\}$. This interference now falsifies the precondition required for the delete operation. If the delete operation had acquired lock on $\{x,\tau\}$, the concurrent insert could have been pre-empted.       

\section{Performing C++ Code Synthesis}
\label{section:cpp}
 A few assumptions necessary for C++ synthesis are stated:
\begin{enumerate}
    \item The traversal code is assumed to be given to \locksynth
    \item The points in which \locksynth synchronization code is to be inserted is clearly specified. 
    \item The basic blocks where destructive update operations are to be performed are clearly annotated, post traversal operations
    \item The notation \texttt{@@<Operation>::<BlockNum>} is used to indicate to \locksynth the basic block where the requisite destructive update (and synchronization) needs to occur. 
    \item The beginning and end of destructive update operation (including traversal) is marked using \texttt{@@begin-destructive-update} and \texttt{@@end-destructive-update} respectively.  
\end{enumerate}
\smallskip\noindent\textbf{Code Synthesis for External Binary Search Trees:}
External binary search trees are similar to binary search trees except that the leaf nodes store the data structure keys. The internal nodes are used for routing purposes. Every internal node should have two children.  For an insert operation, the window of insertion involves the external node that is visited upon BST traversal and its parent. The precondition uses the following ASP relations.
 \texttt{external(x)} : x is an external node,
  \texttt{left(x,y)} : y is the left child of x (x is parent of y),
 \texttt{right(x,y)} : y is the right child of y (x is parent of y),
  \texttt{key(x,kx)} : node x has kx as its key value,
  \texttt{eq(kx,ky)} : keys kx and ky are equal,
  \texttt{lt(kx,ky)} : key kx is less than ky,
  \texttt{reachable(x)} : node x is reachable from the root and 
  \texttt{traversal\_path(x,target)} : node x is on the BST traversal path for the target node

The traversal code for insert operation contains four basic blocks that capture the four different ways in which an insert operation might be performed in an External BST.  Because the analysis by \locksynth is successful in generating a fine-grained synchronization algorithm, the code generation can be performed. \locksynth generates the synchronization code for each of the four blocks and substitutes the block identifiers with the generated code. The precondition and program steps for the first block encoded in ASP are shown. (Table \ref{tab:code_annotation}). Note that the entire C++ code template for insert operation of External BST including the traversal code is captured in the relation \texttt{code/3} on the left in Table \ref{tab:code_annotation}. This relation is essentially an SWI-Prolog fact and will be used as input by \locksynth during the concurrent code generation process. The precondition and program steps for just \texttt{block1} of External BST insert are shown on the right hand side of Table \ref{tab:code_annotation}. They are also SWI-Prolog facts and contain literals from the background theory of External BSTs.

 \begin{table}
 \begin{tabular}{p{7cm}p{5cm}}
 \begin{minted}[frame=single,fontsize=\scriptsize,escapeinside=||]{Prolog}
code(external_bst, insert, '
bool insert(int key){
  struct node * parent = root;
  struct node * curr = root;
  while(curr->left != NULL && curr->right != NULL){
	parent = curr;
	if(curr->key < key)
	   curr = curr->right;
	else
	   curr = curr->left;
    if(parent->left == curr && key < curr->key){
		  @@insert_ext_tree::block1
	 }
    if(parent->left == curr && key >= curr->key){
		@@insert_ext_tree::block2
      }
    if(parent->right == curr  && key < curr->key ){
		  @@insert_ext_tree::block3
	 }
    if(parent->right == curr && key >= curr->key){
            @@insert_ext_tree::block4
     }		
	return false;
}}').
 \end{minted}

 &
 \begin{minted}[fontsize=\scriptsize,frame=single]{prolog}
 pre(external_bst, insert, 
 block1, [ external(y), left(x, y), 
  traversal_path(y, target),
  key(target, ktarget), 
  lt(ktarget, ky),
  not(reachable(target)),
  not(reachable(internal), 
  key(y, ky), 
  key(internal, kinternal), 
  eq(kinternal, ky)
 ]).

program_steps(external_bst, insert, 
block1, [
   link_left(internal, target),
   link_right(internal, y),
   link_left(x, internal)
])
 \end{minted}
 \end{tabular}
 \captionof{table}{Code with annotations (left) and sequential code for External BST (right)}
 \label{tab:code_annotation}
 \end{table}

\smallskip\noindent{\bf Mapping Variables Used by Traversal to Nodes in Declarative Precondition:}
The precondition represented in ASP is declarative and it is not obvious how the traversal code finds a window that matches a particular precondition. Not all the nodes used in the precondition map to the variables used by the traversal code. The traversal code above uses the pointer variables \{parent, curr\} whereas the precondition written in ASP uses the nodes \{x, y, target, internal\}. 
In order to facilitate the synthesis, the variables used by the traversal code are mapped to the nodes used in the ASP precondition. \locksynth will infer the L-values of other nodes that do not have mapping be inspecting the declarative precondition. 
The mapping should be provided by the user as shown (Listing \ref{code:mapping}). 
\begin{code}
\begin{minted}[frame=single, fontsize=\scriptsize]{prolog}
 mapping(external_bst, insert, block1, parent, y).
 mapping(external_bst, insert, block1, curr, x).
 mapping_r_value(external_bst, insert, block1, target->key, key).
 mapping_r_value(external_bst, insert, block2, target->key, key).
 mapping_r_value(external_bst, insert, block3, target->key, key).
 mapping_r_value(external_bst, insert, block4, target->key, key).
\end{minted}
\captionof{listing}{\bf Mapping of l-values and r-values}
\label{code:mapping}
\end{code}

\smallskip\noindent{\bf Memory Allocation:}
 The insert operation for External BST allocates two nodes on the heap before linking them appropriately. \locksynth allocates memory for each node that is declared to be not reachable in the precondition. 
 
\smallskip\noindent{\bf Initialization of Allocated Memory:}
By default, \locksynth assigns the C++ NULL constant to the allocated node properties that represent pointers. If a new node is allocated for a linked list, \locksynth initializes the \texttt{next} property to NULL. For binary trees, \locksynth initializes both \texttt{left} and \texttt{right} pointers to NULL. 

\smallskip\noindent{\bf Inferring L-values of Precondition Nodes:}
  The nodes referenced in the declarative precondition are related to one another through either the left/2 or right/2 relations. In case of Linked Lists, they are related to one another through the edge/2 relation. Structurally, left(x, y) signifies that x is the predecessor of y in the heap in the pre-order traversal of the tree. Because the data structures considered in this paper are trees, this convention of a predecessor node is followed. Similarly, right(x,y) signifies that x is the predecessor of y and likewise for edge(x,y). Therefore, if x has already an assigned L-value (from the traversal code mapping), then y's L-value can be inferred using this convention. If left(x, y) is true in the precondition and x is mapped to curr, then y's L-value is \texttt{curr->left}. \locksynth uses the L-value of the immediate predecessor. Further, the L-value of the immediate predecessor is computed recursively. The relation edge/2 is used for \texttt{next} pointer. 

\begin{table}
 \begin{tabular}{p{6cm}p{6cm}}
 \begin{minted}[frame=single,fontsize=\scriptsize,escapeinside=||]{Prolog}
infer_l_value(Op, Block, Node, LValue) :-
   ds(DataStructure),
   mapping(DS, Op, Block, LValue, Node), !.      

infer_l_value(Op,Block,
    Node,LValue->Relation) :-
   ds(DataStructure),
   precondition(DS, Op, Block, Precondition),
   immediate_predecessor(Node, 
             Precondition, Pred, Relation),
   infer_l_value(Op, Block, 
           Predecessor, LValue), !.

immediate_predecessor(Node,Pre,Pred,next) :-
      member(edge(Pred,Node),Pre).
immediate_predecessor(Node,Pre,Pred,left) :-
      member(left(Pred,Node),Pre).
immediate_predecessor(Node,Pre,Pred,right) :-
      member(right(Pred,Node),Pre).
 \end{minted}
 &
 \begin{minted}[fontsize=\scriptsize,frame=single]{prolog}
infer_r_value(Op, Block, Node,
          Prop, RValue) :-
  ds(DataStructure),
  precondition(DataStructure,Op,Block,Pre),
 infer_l_value(Op, Block, Node, LValue),
 mapping_r_value(DataStructure, Op,
        Block, LValue,Prop,RValue).

infer_r_value(Op, Block,
       Node, Prop, LValue->Prop) :-
 ds(DataStructure),
 pre(DataStructure, Op, Block, Pre),
 infer_l_value(Op,Block,Node, LValue),
 LValue \= Node.
\end{minted}
 \end{tabular}
 \captionof{table}{Prolog code within \locksynth to infer l-values and r-values}
 \label{tab:infer}
 \end{table}

\smallskip\noindent{\bf Inferring R-values of Precondition Node Properties that are Newly Allocated:}
 For already allocated nodes that are part of traversal, R-values need not be inferred. However, for newly allocated nodes, the R-values of node properties should be inferred. In the case of external BST insert operation, the R-value for target node key is the key that is going to be inserted into the tree. Further, the R-value of internal node key is the key of the external node in the traversal. \locksynth uses the eq/2 relation to infer that the R-value of internal node's key is that of the last external node traversed. Because the target key value is served as an input to the insert operation, its mapping must be made explicit as shown (Listing \ref{code:mapping}, Table \ref{tab:infer}).

\smallskip\noindent{\bf Inferring Correct Locking and Unlocking Order:}
 To avoid deadlocks, every thread in a concurrent execution should acquire and release locks in a uniform order. Therefore, \locksynth acquires locks in the order of predecessors of nodes referenced in the declarative precondition. \locksynth starts by selecting node that does not have a predecessor and then selects the successors in a breadth-first search. This will be the same order in which locks on the selected nodes will be acquired and released, which guarantees deadlock freedom. 
 
 \smallskip\noindent{\bf Inserting Validation Condition:}
   As stated before, precondition validation logic code for each block is assumed. The \texttt{validate/4} fact captures the precise precondition validation for a given data structure's operation and block. An example for External BST is provided in Listing \ref{code:validatefacts}.   
   
\small\noindent{\bf Performing Read-Copy-Update (RCU) Synthesis for Internal Binary Search Trees:}
 In order to perform RCU synchronization, the beginning and end of traversal code need to be clearly marked by the user using \texttt{@@begin-traversal} and \texttt{@@end-traversal}. These annotations will be respectively replaced by \texttt{rcu\_read\_lock()} and \texttt{rcu\_read\_unlock()} statements. Further, just before performing destructive update steps, \locksynth inserts the \texttt{rcu\_synchronize()} statement. The RCU-Synchronize statement blocks until all traversals are done and then synchronizes the destructive updates back to the tree atomically \cite{mckenney2013rcu}.  

\begin{code}
  \begin{minted}[frame=single, fontsize=\scriptsize]{prolog}
validate(external_bst, insert, block1, 'reachable(parent) 
&& parent->left == curr && curr->left == NULL && curr->right == NULL').
validate(external_bst, insert, block2, 'reachable(parent) 
&& parent->left == curr && curr->left == NULL && curr->right == NULL').
validate(external_bst, insert, block3, 'reachable(parent) 
&& parent->right == curr && curr->left == NULL && curr->right == NULL').
validate(external_bst, insert, block4, 'reachable(parent) 
&& parent->right == curr && curr->left == NULL && curr->right == NULL').
   \end{minted}
   \captionof{listing}{Preconditon validation logic captured as \texttt{validate/4} facts}
\label{code:validatefacts}
\end{code}

\begin{code}
\begin{minted}[frame=single,fontsize=\scriptsize]{cpp}
//Allocation of unreachable nodes
struct node * internal = (struct node *) malloc(sizeof(struct node));
struct node * target = (struct node *) malloc(sizeof(struct node));
//Initializing pointers to NULL values for allocated nodes
internal->left = NULL; internal->right = NULL; target->left = NULL; target->right = NULL;
//Setting inferred R-Values for Keys
internal->key = curr->key; target->key = key;
//Lock order in the order of predecessors
parent->mtx.lock(); target->mtx.lock(); internal->mtx.lock(); curr->mtx.lock();
//Validation logic for Optimistic Synchronization
if(!(reachable(parent) && parent->left == curr && curr->left == NULL && curr->right == NULL)){
//Locks released in the same order of lock acquisition
parent->mtx.unlock(); target->mtx.unlock(); internal->mtx.unlock();
curr->mtx.unlock();
continue;  // Note that insert operation will retry
}
// Destructive update steps re-ordered and L-Values inferred by Locksynth
internal->left = target; internal->right = curr; parent->left = internal;
//Same as unlocks statements before
parent->mtx.unlock(); target->mtx.unlock(); internal->mtx.unlock();
curr->mtx.unlock();
\end{minted}
\captionof{listing}{{\bf Generated Code for an Insert Operation Basic Block}}
\label{code:cppcodegen}
\end{code}

\section{Performance Comparison of Synthesized Code with Hand-crafted Code}
\label{section:synchrobench}
The Synchrobench microbenchmark compares the performance of several concurrent algorithms from literature \cite{gramoli2015more}. The benchmark consists of C/C++ and Java implementations of Linked Lists, Binary Search Trees and Balanced Trees. Because \locksynth can only generate the synchronization code for insert and delete operations for Linked Lists, External BSTs and Internal BSTs, their equivalent hand-crafted counterparts from the benchmark are compared. The data structures are tested for key size ranges starting from 100 till 10 million multiplying by a factor of 10 at each range. The initial size of the dictionary is set to 50\% of the key range. Further, the workloads considered are read- dominated workloads (100\% reads and 0\% updates), write-dominated workloads (100\% writes split evenly between insert and delete and 0\% reads) and mixed workloads (70\% reads, 30\% updates split evenly between insert and delete). All experiments were run on an Intel(R) Xeon(R) Gold 5220R processor @ 2.20 GHz with 48 logical cores and 247 GB RAM.   
All the programs were compiled using g++ compiler with level 3 optimization. For hand-crafted linked lists, we use the
C++ version of the lazy marking algorithm \cite{herlihy2020art}.  We did not find a lock-based external BST implementation in the benchmark. Therefore we used Natarajan and Mittal's lock-free external binary search tree. Natarajan and Mittal's tree locks edges in the tree and performs some book-keeping to manage concurrent deletes \cite{natarajan2014fast}. 
For internal BST, the hand-crafted version is the Citrus RCU tree \cite{arbel2014concurrent}. The effective update rate is the percentage of update operations that actually made changes to the data structure among all submitted operations. The effective update ratios of \locksynth version to the hand-crafted versions are also presented. A ratio greater than 1 for throughput indicates \locksynth outperformed the hand-crafted version. Likewise, a higher update rate ratio that the algorithm performs more destructive updates in the presence of contention. Because optimistic algorithms are better for read-dominated workloads, the effective update rate generally is poor. The \locksynth Linked List outperforms the hand-crafted lazy list when comparing the effective update ratio for all workloads. However, the write-throughput is 7 times worse (not shown) when the key sizes increase. 
The \locksynth External BST performs very well for read-dominated workloads, when considering the fact the hand-crafted External BST is a lock-free version. As the key ranges increase, the throughput slightly degrades to a factor less than one-third of the lock-free version. For write-dominated workloads and mixed workloads, the throughput of \locksynth is roughly 0.25 times the throughput of lock-free version. This is due to high contention at the external nodes of the External BST and lock-based versions are expected to perform poorly. The effective update rate is surprisingly high for key range of 100. Then, the update ratio goes to 0.3 till 100,000 keys. From a million onwards, the upate ratio again increases and is greater than 1 (about 8-9 times greater) for 10 million than the lock-free version. This behavior is same for both mixed and write-dominated workloads. This might be due to additional memory reclamation logic that the lock-free BST performs. The throughput and effective update rate behavior for the mixed workload remains very similar to the write-dominated workload.
The \locksynth Internal BST performs better than the hand-crafted Citrus tree for read-dominated workloads. This can be explained by the fact that there is a lot of book-keeping done by Citrus tree that aids write-dominated and mixed workloads. For write-dominated workloads, the throughput for \locksynth BST is very good. However, the update rate drops to near zero relative to what Citrus tree does. This indicates that Internal BST is extremely slow at making changes when there is high contention and most of the update operations fail to make any changes to the data structure. This is where the hand-crafted Citrus tree and in general any algorithm that is designed with a lot of book-keeping can perform better. The performance behavior for mixed workload is similar to the write-dominated workload. 

\begin{figure}[tb]
    \begin{tikzpicture}[scale=0.7, transform shape]
        \begin{groupplot}[group style={group size = 2 by 1}, 
            height=2.0in, xtick=data,
            /pgf/bar width=7.5pt,
            legend style={at={(-0.5,1.2)}, anchor=south, legend columns=-1, /tikz/every even column/.append style={column sep=10pt}}]

            \nextgroupplot[ybar, width=4.5in, symbolic x coords = {LL-MW,LL-WD,BST-MW,BST-WD}, enlarge x limits=0.25, ylabel={Throughput Ratio}]
            \addplot coordinates {
                (LL-MW, 1) (LL-WD, 1.288) (BST-MW, 0) (BST-WD, 0)
            };
            \addplot coordinates {
                (LL-MW, 1) (LL-WD, 1.018) (BST-MW, 0.002) (BST-WD, 0.002)
            };
            \addplot coordinates {
                (LL-MW, 1) (LL-WD, 1.003) (BST-MW, 0) (BST-WD, 0.001)
            };
            \addplot coordinates {
                (LL-MW, 1) (LL-WD, 1.045) (BST-MW, 0) (BST-WD, 0.001)
            };
            \addplot coordinates {
                (BST-MW, 0) (BST-WD, 0.001)
            };
            \addplot coordinates {
                (BST-MW, 0) (BST-WD, 0.001)
            };
 
            \nextgroupplot[ybar, width=3.0in, symbolic x coords = {EBST-MW,EBST-WD}, enlarge x limits=0.5]
            
            \addplot coordinates {
                (EBST-MW, 1777.642) (EBST-WD, 2278.483)
            };
            \addplot coordinates {
                (EBST-MW, 277.967) (EBST-WD, 18.418)
            };
            \addplot coordinates {
                (EBST-MW, 1.719) (EBST-WD, 0.0011)
            };
            \addplot coordinates {
                (EBST-MW, 0.473) (EBST-WD, 0.32072)
            };
            \addplot coordinates {
                (EBST-MW, 0.818) (EBST-WD, 0.596)
            };
            \addplot coordinates {
                (EBST-MW, 7.897) (EBST-WD, 9.108)
            };
           \legend{100,1K,10K,100K,1M,10M} 
        \end{groupplot}
       
    \end{tikzpicture}
\caption{Comparison of effective update ratios of synthesized data structures under mixed (MW) and write-dominated (WD) workloads for different key space sizes.}
\label{fig:my_label}
\end{figure}
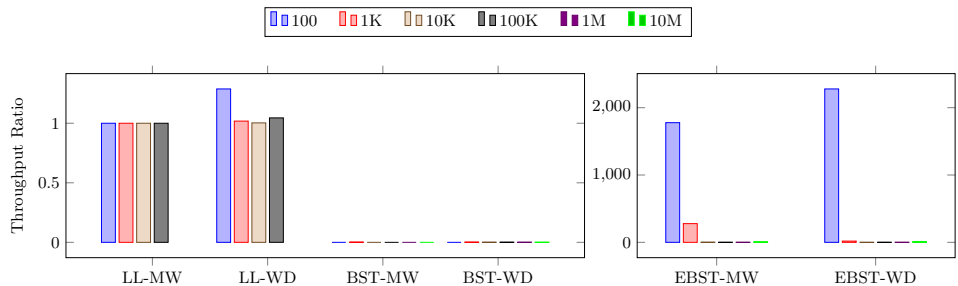

\remove
{
\begin{figure}
    \centering
    \includegraphics[scale=0.7]{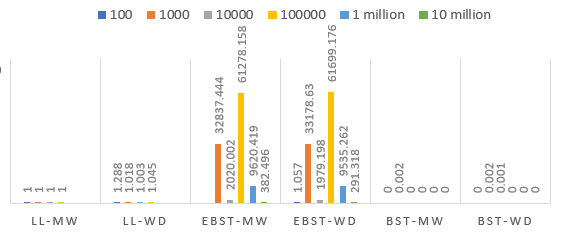}
    \caption{Comparison of Effective Update Ratios of Synthesized Data Structures under Mixed workload (MW) and Write-dominated workloads}
    \label{fig:my_label}
\end{figure}
}

\section{Related Work}
\label{section:related}
One of the first works in concurrency synthesis is the seminal work of Amir Pnueli et al. towards synthesis of synchronization skeletons \cite{clarke1981design}. They use branching time temporal logic, on the state space of programs, to model mutual exclusion of critical sections and starvation freedom. Essentially model checking on CTL formulas is performed through clear identification of critical sections. There are several works that have performed verification of concurrent data structures by varied assumptions.
Then, Vafeiadis et al. established Rely-Guarantee (RG) reasoning for fine-grained concurrent data structures \cite{vafeiadis2006proving}. Their proof method is an extension of Owicki Gries \cite{owicki1975proving} reasoning originally formulated to reason about parallel programs. 
\par \noindent The rely-guarantee reasoning by Vafeaidis et al. relies on identification of linearization points to reason about correctness. Our method does not rely on explicit identification of linearization points. Further, Vafeaidis et al. combined Rely-Guarantee with well known Separation Logic (SL) into a new logic called \textit{RGSep} \cite{vafeiadis2007marriage}. \textit{RGSep} can perform verification of lock-coupling and lock-free concurrent lists. A key step in the proof relies on stabilizing the post-conditions of each thread without which Rely-guarantee proof cannot be completed. To stabilize the post-conditions of a thread, the proof technique should infer valid \textit{frames} in the heap that are not touched by every thread. However, the technique does not perform additional reasoning on the heap unlike our knowledge-guided approach. For instance, that RCU is required on the Tree based structure due to missed traversals cannot be inferred using \textit{RGSep}, where the proof of correctness would simply report an error or fail to terminate.  \textit{RGSep} has been mechanized and due to relatively low involvement of arithmetic operations, the tool has achieved considerable success. However, our approach based on ASP can easily express arithmetic constraints (CLP(Q)) such as height-balancing of trees and precisely compute whether a set of locks are adequate.  Vechev et. al \cite{vechev2010abstraction} have used bounded model checkers to (semi-)automatically generate lazy list synchronization algorithms. Their technique too relies on identification of linearization points. They have to provide additional meta-data in order to generate more sophisticated algorithms. This is similar to using knowledge-based approaches to synthesis of synchronization.
\par\noindent
 Very recently, Occualizer framework \cite{shanny2022occualizer} generates optimisitic RCU based synchronization code for tree data structures from a given sequential input version. Occualizer work makes certain assumptions about the input tree data structure and the assumptions are manually checked by source code inspection. This approach can be automated. Occualizer can handle more complex trees such as External Red-Black trees, B+ trees and Radix trees that Locksynth cannot. This is because Locksynth cannot handle the book-keeping logic involved in performing a variable number balancing operations that might terminate at the root node of a tree in general.  Nonetheless, Occualizer can only generate an RCU algorithm even for a Linked List and External Binary Search Tree based on sequential source code inspection. However, Locksynth can generate the optimistic version while acquiring the precises number of locks sufficient to perform the concurrent updates.  Also, as part of future work, due to the Knowledge representation and reasoning approach taken by Locksynth, it can incorporate the reasoning involved in Occualizer.

 \section{Conclusion and Future Work}
 \label{section:conclusion}
     \locksynth synthesizes concurrent code for Linked Lists, External BSTs, Internal BSTs and External Balanced Search Trees. It can synthesize concurrent versions of insert, delete for Linked Lists and External BSTs. \locksynth can also recommend RCU framework for Internal BSTs due to key-movement missed by an asynchronous observer. For further illustration, use of \locksynth on External Red-Black trees is shown in Appendix A. Table \ref{tab:results} gives a summary of data structures \& their operations for which \locksynth can automatically synthesize concurrent code. \locksynth can check lock adequacy for tasks such as left/right rotations for height-balanced trees, but cannot yet generate full code as the insert and delete operations require a composition of the smaller tasks such as rotations and weight-adjustments. It is part of our future work. 

     \par Currently, the usability of \locksynth involves a steep learning curve to achieve synthesis. Similar learning curves exist with using model checkers for concurrent programs \cite{kroening2014cbmc} and theorem-prover tools based on concurrent separation logic \cite{brookes2016concurrent, mulder2022diaframe, krebbers2017interactive}. Nonetheless, the formal method toolchains mentioned are mature enough and much easier to use than \locksynth in its current state. All the theorem-prover tools assume an abstract representation of concurrent code and prove correctness over the abstract representation. Whereas, we start with an abstract representation of data structure knowledge and perform actual code generation.   The code generated by \locksynth can be used in conjunction with other model checkers and concurrent separation logic theorem provers to provide an independent path of verification for the generated code.  Improving the usability of our tool and integrating our abstract representations of sequential code with Dafny notations \cite{leino2010dafny} is part of our future work. Further, the advantage of using such abstract representations can make \locksynth generate concurrent Java programs. While performing these translations, some caution should be exercised when considering the memory model guarantees offered by the language run-time specification \cite{manson2005java, boehm2008foundations}. We again leave all this for future work. 
     \par \locksynth represents the first step towards commonsense reasoning approaches to derive concurrent data structures. More work needs to be done to be able to perform data structure traversal synthesis and complete the synthesis for insert and delete operation for External Height-balanced trees considered in this paper. More knowledge can be provided to enable the derivation of much faster concurrent algorithms. Relevant work by Occualizer hints at making some assumptions on the data structure needed to match performance of hand-crafted versions stronger \cite{shanny2022occualizer}. We also aim to support more sophisticated atomic instructions  \textit{Compare-and-Swap} \cite{valois1995lock}, \textit{Fetch-and-Add} \cite{heidelberger1990parallel}, offered by modern multi-processors. Towards that end, more concurrency expert knowledge needs to be captured to facilitate the derivation of lock-free data structures. Finally, goal-directed ASP using s(CASP) \cite{arias2018constraint} can be used to perform the backend reasoning instead of Clingo. However, s(CASP) still requires effective dynamic consistency checking of constraints \cite{arias2022towards} and tabling \cite{ariasthesis} to be performant on dynamic domains, although some relevant work has been done using the Event calculus formalism \cite{arias2022modeling, varanasi2022modeling}.   
       

 \begin{table}
 \centering
 \scriptsize
 \begin{tabular}{p{2cm}|p{3cm}|p{1cm}|p{1cm}|c|p{1cm}}
  \cline{1-6} 
 Data Structure & Operations & Lock Adequacy & Program Order  & Synthesized Algorithm & Code gen
 \\
 \cline{1-6} 
 Linked List & 
      insert, delete
  &  \centering Yes  & \centering Yes  & Fine-grained & Yes\\
 \cline{1-6} 
 External BST & 
     insert, delete
   & \centering Yes   & \centering Yes  & Fine-grained & Yes\\
 \cline{1-6} 
 Internal BST & 
     insert, delete
   & \centering Yes  & \centering No  & RCU & Yes \\ 
 \cline{1-6} 
 External AVL &
   add,remove,
   left/right rotate
   & \centering Yes  & \centering No  & Coarse-grained/RCU & No \\
 \cline{1-6} 
 External RB & 
    add, remove,
    red-red-conflict,
   left/right rotate 
  &  \centering Yes   & \centering  No  & Coarse-grained/RCU & No \\
 \cline{1-6} 
 Chromatic Search tree & 
  add,remove,
  single/double rotation,
  adjust weights,
  rotate-adjust-weight
   & \centering Yes  & \centering No  & Coarse-grained/RCU & No 
 \\
  \cline{1-6}

 \end{tabular}
 \captionof{table}{\locksynth results on Linked Lists and (External) Binary Search Trees}
 \label{tab:results}
 \end{table}

\smallskip 
\noindent{\bf Acknowledgement:} We thank the anonymous reviewers for their insightful comments.
 
\printbibliography

\end{document}